\documentclass[prl,twocolumn,groupedaddress,nofootinbib]{revtex4}

\def\slashchar#1{\setbox0=\hbox{$#1$}
   \dimen0=\wd0 \setbox1=\hbox{/} \dimen1=\wd1
   \ifdim\dimen0>\dimen1 \rlap{\hbox to \dimen0{\hfil/\hfil}} #1
   \else  \rlap{\hbox to \dimen1{\hfil$#1$\hfil}} / \fi}

\def\({{\Bigl(}}
\def\){\Bigr)}
\def\]{\Bigr]}
\def\[{\Bigl[}

\begin{document}

\title{Comment on ``A quantum-classical bracket that satisfies the Jacobi identity''
[J. Chem. Phys. {\bf 124}, 201104 (2006)]}

\author{L. L. Salcedo}
\email{salcedo@ugr.es}

\affiliation{
Department of Atomic, Molecular, and Nuclear Physics,
University of Granada,
E-18071 Granada, Spain}

\date{\today} 



\maketitle

The quantum mechanical description of microscopic systems is nowadays
well established and cleanly formulated (at least above the very small
Planck scale, where quantum gravity awaits a firmer foundation).
However, in many cases, even if the wave equations to be solved are
known, they are not easily amenable to a full quantum computation, in
practice. A standard approach is then to resort to approximate
descriptions of the semiclassical type, where some of the degrees of
freedom of the full system are treated quantum mechanically while
others are treated at a classical level. In this approach one ends up
with a mixed quantum-classical system.

The canonical structures of both classical and quantum mechanics are
based on the existence of a Lie bracket between observables, $(A,B)$;
this is the commutator in the quantum case and the Poisson bracket in
the classical case.  The dynamical bracket is such that if $G$ is the
infinitesimal generator a symmetry transformation, the variation of an
observable $A$ takes the following form
\begin{equation}
\delta A= (A,G) \,\delta\lambda \,.
\end{equation}
The Lie bracket properties
\begin{eqnarray}
&& (a A+bB,C)=a(A,C)+b(B,C) \quad \text{(linearity),} \nonumber\\ &&
(A,B)=-(B,A) \quad \text{(antisymmetry),} \\ && ((A,B),C)+ ((B,C),A) +
((C,A),B) = 0 \quad \text{(Jacobi),} \nonumber
\end{eqnarray}
encapsulate the group structure of symmetry transformations, including
the dynamical evolution, 
\begin{equation}
\frac{d}{dt}A= (A,H)+\frac{\partial}{\partial t}A
\end{equation}
($H$ being the Hamiltonian), and are therefore crucial for a fully
consistent physical description of the classical or quantum system.

In view of this similarity between the classical and quantum
formulations, many authors have considered the possibility of finding
a consistent description for the abovementioned mixed
quantum-classical systems, i.e., systems composed of two interacting
sectors, a quantum one and a classical one. Note that such mixed
systems are perfectly legitimate as {\em approximations} to an exact
quantum-quantum system, the challenge is, however, to find an {\em
autonomous} and closed description for the mixed case, rather than an
intrinsically approximated one. As is well known, ad hoc
approximations tend to break exact properties (e.g., symmetries,
conservations laws, unitarity, etc) of a theory, while internally
consistent approximations tend to preserve them, and so they are, in
principle, preferable. (Classical mechanics is precisely an example of
a consistent approximation, namely, to quantum mechanics.)

A simple case, often considered in the literature, is that of a one
dimensional system with two structureless particles, where $x$ and $k$
are the position and momentum variables of the classical particle,
respectively, and $q$ and $p$ those of the quantum particle. $x$ and
$k$ commute with everything while $[q,p]=i\hbar$. In this case, the
observables are constructed with $x$, $k$, $q$ and $p$,
\begin{equation}
A=\sum_{n,m,r,t}a_{nmrt}\,x^n k^m q^r p^t \,.
\end{equation}

Minimal requirements for a consistent quantum-classical formulation
would include the following:
\begin{itemize}
\item[(i)] Symmetry transformations are carried out by a dynamical
bracket between observables that must be a Lie bracket.
\item[(ii)] The dynamical bracket between two purely quantum
observables should reduce to the standard one,
\begin{equation}
(Q,Q^\prime)=\frac{1}{i\hbar}[Q,Q^\prime]=\frac{1}{i\hbar}(QQ^\prime-Q^\prime Q) \,,
\end{equation}
and likewise, for two purely classical observables
\begin{equation}
(C,C^\prime)=\{C,C^\prime\}=
\frac{\partial C}{\partial x}\frac{\partial C^\prime}{\partial k}
-\frac{\partial C}{\partial k} \frac{\partial C^\prime}{\partial x}
\,.
\end{equation}
\end{itemize}

A standard proposal, made independently by several authors
\cite{Aleksandrov:1981,Boucher:1988ua}, is as follows,
\begin{equation}
(A,B)= \frac{1}{i\hbar}[A,B] +\frac{1}{2}( \{A,B\}- \{B,A\})
\label{eq:7}
\,.
\end{equation}
Unfortunately, this definition fails to satisfy the Jacobi
identity. An explicit counterexample is provided in \cite{Caro:1998us}, for
$A=xq$, $B=xqp$, and $C=k^2 p$, since
\begin{equation}
((A,B),C)+ ((B,C),A) + ((C,A),B) =  \frac{1}{2}\hbar^2 \,.
\end{equation}
(Jacobi identity violations always come from finiteness of $\hbar$,
since in the classical limit any bracket must revert to the Poisson
bracket which does satisfy Jacobi.)

In a recent work \cite{Prezhdo:2006}, a new proposal is made, claiming to define
a true Lie bracket. No mathematical proof is provided, but it is shown
(correctly) that Jacobi is satisfied for the three observables of the
example just discussed.  Regrettably, as I show below, the Jacobi
identity is not satisfied by this new bracket either, if one takes
three generic observables.

In the new proposal of \cite{Prezhdo:2006}
\begin{equation}
(A,B)= (A,B)_q+ (A,B)_c
\,,
\label{eq:9}
\end{equation}
with
\begin{equation}
(A,B)_q= \frac{1}{i\hbar}[A,B] \,,
\label{eq:10}
\end{equation}
as in (\ref{eq:7}). However, the classical part differs from
(\ref{eq:7}). The new prescription is to take the Poisson bracket of
the classical variables involved, while the quantum variables are
ordered by moving (commuting) the $q$'s to the left of the $p$'s and
setting the $\hbar$ so generated to zero. Of course, this is
equivalent to a ``normal order'' prescription in which $q$ is set to
the left of $p$ by hand, $:pq:=qp$. (For simplicity, I use the
standard notation $:~:$ to denote this ``normal order'' of operators.)
That is,\footnote{More precisely,
$$
(:{\cal A}(x,k,q,p):,:{\cal B}(x,k,q,p):)_c= :\{{\cal A}(x,k,q,p),{\cal B}(x,k,q,p)\}: \,.
$$}
\begin{equation}
(A,B)_c= \; :\{A,B\}: \,,
\end{equation}
where the Poisson bracket affects the classical variables. More
explicitly, for two observables,
\begin{eqnarray}
A &=&\sum_{n,m}a_{nm}(x,k)\,q^np^m \,, \nonumber \\ 
B
&=&\sum_{r,t}b_{rt}(x,k)\,q^r p^t \,,
\end{eqnarray}
$a_{nm}(x,k)$ and $b_{rt}(x,k)$ being ordinary functions on the phase
space of the classical particle,
\begin{eqnarray}
(A,B)_c=
\sum_{n,m}\sum_{r,t}\big\{a_{nm}(x,k),b_{r t}(x,k)\big\}\,
q^{n+r} p^{m+t} \,.
\label{eq:13}
\end{eqnarray}
As I said, this definition does not preserve the Jacobi identity. An
explicit counterexample is provided by the new triple $A=kp$, $B=xp$,
and $C=q^2$. For these observables, one easily finds
\begin{eqnarray}
(A,B) &=& -p^2 \,, \quad ((A,B),C)=4qp-2i\hbar, \nonumber \\
(B,C) &=& -2xq \,, \quad ((B,C),A)=-2xk-2qp,  \nonumber \\
(C,A) &=& 2kq \,, \quad ((C,A),B)=2xk-2qp,  \nonumber 
\end{eqnarray}
and hence,
$$
((A,B),C)+ ((B,C),A) + ((C,A),B) = -2i\hbar \,.
$$ Therefore the Jacobi identity is violated. (However, Jacobi is
preserved by this triple with the original bracket (\ref{eq:7}). The
same is true whenever the observables involved are at most quadratic
in the dynamical variables $x$, $k$, $q$, and $p$, \cite{Caro:1998us}.)

Such violation of the Lie bracket property is not surprising; in
\cite{Caro:1998us} a no-go theorem was proven (see also
\cite{Terno:2004ti,Sahoo:2004}), namely, if one requires the
quantum-classical bracket to fulfill the rather natural axioms
\begin{equation}
(CQ,C^\prime)=\{C,C^\prime\} Q \,,\quad
(CQ,Q^\prime)=\frac{1}{i\hbar}[Q,Q^\prime] C \,,
\end{equation}
($C$, and $C^\prime$ being purely classical and $Q$ and $Q^\prime$ purely
quantum observables), then Jacobi cannot be satisfied. (Note that the
bracket (\ref{eq:9}), with (\ref{eq:10}) and (\ref{eq:13}), satisfies
the axioms.)

Finally, let me note that in addition to the requirements (i) and
(ii), another natural property (common to classical and quantum
mechanics) is that the product of two observables $AB$ at $t=0$ should
evolve into the product $A(t)B(t)$ and time $t$ (and similarly for
other symmetry transformations). This implies the 
\begin{itemize}
\item[(iii)] Leibniz rule property for the dynamical bracket,
\begin{equation}
(AB,C)= (A,C)B+A(B,C)\,.
\end{equation}
\end{itemize}
As shown in \cite{Salcedo:1996jr}, a bracket fulfilling (i-iii) is necessarily
either the Poisson bracket (no quantum sector) or the quantum
commutator (no classical sector). No quantum-classical mixture is
allowed.

In summary, the proposal in \cite{Prezhdo:2006} does not actually define a Lie
bracket since it fails to satisfy the Jacobi identity. Furthermore,
any quantum-classical bracket must have awkward properties, as it has
to violate rather natural requirements, satisfied both in the purely
classical or purely quantum cases. This just means that the
quantum-classical mixing remains as a useful but intrinsically
approximated approach.

\begin{acknowledgments}
This work was supported by DGI, FEDER, UE, and Junta de Andaluc{\'\i}a funds 
(FIS2005-00810, HPRN-CT-2002-00311, FQM225).
\end{acknowledgments}


\end{document}